\documentclass[]{aa}
\usepackage{graphicx}
\usepackage{txfonts}
\usepackage{natbib}

\begin{document}
\title{Collisions between equal sized ice grain agglomerates}

\author{C.~Sch{\"a}fer \and R.~Speith \and W.~Kley}
\offprints{R.~Speith}
\institute{Institute for Astronomy \& Astrophysics, University of
T{\"u}bingen, Auf der Morgenstelle 10, D-72076 T{\"u}bingen.\\ 
\email{\{schaefer,speith,kley\}@tat.physik.uni-tuebingen.de} 
}
\date{Received ---; accepted ---}
 
\abstract
{Following the recent insight in the material structure of comets,
protoplanetesimals are assumed to have low densities and to be highly
porous agglomerates. It is still unclear if planetesimals can be formed
from these objects by collisional growth.}
{Therefore, it is important to study numerically the
  collisional outcome from low velocity impacts of equal sized
  porous agglomerates which are too large to be examined in a
  laboratory experiment.}
{We use the Lagrangian particle method Smooth Particle Hydrodynamics to
solve the equations that describe the dynamics of
elastic and plastic bodies. Additionally, to account for the influence
of porosity, we follow a previous developed equation of state and
certain relations between the material strength and the relative
density.}
{Collisional growth seems possible for rather low collision velocities
and particular material strengths. The remnants of collisions with
impact parameters that are larger than 50~\% of the radius of the
colliding objects tend to rotate. For small impact parameters, the
colliding objects are effectively slowed down without a prominent
compaction of the porous structure, which probably increases the
possibility for growth. The protoplanetesimals, however, do not stick
together for the most part of the employed material strengths.}
{An important issue in subsequent studies has to be the influence of
  rotation to collisional growth. Moreover, for realistic
  simulations of protoplanetesimals it is crucial to know the
  correct material parameters in more detail.}


\maketitle

\section{Introduction}

The initial growth of particles in a protoplanetary disc is
accomplished by sticking collisions. All solid objects in a planetary
system are believed to have been developed from collisions between
small dust particles with initial sizes of about $0.1$ micron
\citep{1982come.coll..131G}. These collisions result from Brownian
motion, gas turbulence in the disc and gas drag
\citep{1977MNRAS.180...57W,1980A&A....85..316V,1993prpl.conf.1031W}. It
has been shown that even protoplanetesimals of equal mass might have
rather high relative velocities due to different shapes and hence
different gas resistances \citep{2000SSRv...92..279B}, depending on
the orbital distance to the protostar. Experimental data for
collisional growth of macroscopic agglomerates with sizes extending
the dm-regime are rather scarce. The formation of such macroscopic
bodies from smaller dust grains was studied in detail by
\cite{1993A&A...280..617O} and the fractal growth was modelled by
\cite{1997ApJ...480..647D} and \cite{1999Icar..141..388K}. Generally,
the dust particles stick together due to van-der-Waals forces. This
growth process leads to (highly) porous agglomerates, a fact that was
experimentally confirmed by \cite{1998Icar..132..125W}. These
macroscopic agglomerates do not stick together as easily as tiny dust
grains which only interact by van-der-Waals forces. Instead, aggregate
compaction, fragmentation and disruption become important above a
specific kinetic energy of the collision. The agglomerates can break
and can be dispersed which eventually prevents a fast growth process
or even further growth at all. It is now generally assumed that
decimetre sized porous agglomerates can form rather quickly in the
presolar nebula and the protoplanetary accretion disc
\citep{2000ApJ...533..454P,1998Icar..132..125W,2000Icar..143..138B,2000SSRv...92..295W}.
The critical bottleneck in the formation process of planetesimals lies
however in the size range from about several decimetre to objects
whose interaction is mainly dominated by gravitation, that is 1 to
10~$\mathrm{km}$ planetesimals. First studies to the collisional
growth of brittle planetesimals with material strength were performed
by \cite{1994Icar..107...98B,benz:1995,1999Icar..142....5B} and
\cite{2000SSRv...92..279B}. Their results indicate that the weakest
bodies during collisions are in the size range from metres to around
100~$\mathrm{m}$. Caused by the stronger interaction with the gas,
metre sized bodies experience higher collision velocities and are also
rather fragile objects and easily disrupted.

A better understood problem is the formation of planets from
planetesimals. The formation of planetary embryos or cores has been
investigated in a series of publications
\citep{1988Icar...74..542S,1989Icar...77..330W,1993Icar..106..190W}
with the application of a statistical approach to describe the
evolution of a swarm of planetesimals around the star. Later, by the
use of modern supercomputers, direct n-body simulations of a large
number of planetesimals became feasible and yielded similar results
\citep{2000Icar..143...15K,2002ApJ...581..666K}. Once planetary cores
have formed, the terrestrial planets grow through collisions between
them
\citep{1998Icar..136..304C,1999Icar..142..219A,2002Icar..157...43K}.
In this size regime, the collisional outcome is by far dominated by
gravitational interactions \citep[e.g.,][]{2004ApJ...613L.157A}.

One alternative to avoid the bottleneck at the metre length scale of
the formation process of planetesimals is gravitational instability of
the dust layer in the midplane of the protoplanetary accretion
disc. This idea was suggested independently by
\cite{1969QB981.S26......} and \cite{1973ApJ...183.1051G}. For the
solar nebula, \cite{1973ApJ...183.1051G} found gravitational clumping
of the dust which forms clumps with radii of 10~$\mathrm{km}$ at
1~$\mathrm{AU}$ within 1000~years after the onset of the vertical
settling of dust. However, it is still unclear and strongly discussed
in the community whether the dust in the midplane can become
gravitational unstable, because Kelvin-Helmholtz instability and
turbulent motion of dust particles in the disc lead to significant
velocity dispersions which may prevent the collapse
\citep{1993Icar..106..102C,2000SSRv...92..295W,2002ApJ...580..494Y,2004ApJ...601.1109Y}.

Therefore, collisional growth of metre sized bodies has to be studied
in greater detail in order to investigate the mechanisms of
planetesimal formation. Laboratory experiments of collisions
between water ice objects \citep[e.g.,][]{1996Icar..123..422B} have
been restricted to smaller sizes (up to several $\mathrm{cm}$) and low
impact velocities (up to several $\mathrm{cm/s}$). Thus, experiments
have concentrated on the dependence of sticking forces from the
frost-coating of the surfaces \citep{1997Icar..129..539S} or on
measuring restitution coefficients \citep{1995Icar..113..188S}.

Since real collisions between objects of metre size are not feasible
yet in laboratory experiments, one has to resort to numerical
simulations. Here, we employ the numerical Lagrangian particle method
Smooth Particle Hydrodynamics (SPH) which was first introduced by
\cite{lucy:1977} and \cite{gingold:1977} for modelling compressible
flows in astrophysical problems. The method was extended to solid
state mechanics in the beginning of the nineties by
\cite{libersky:1990} and improved extensively in the following years
\citep{libersky:1993,randles:1996,libersky:1997}. After
\cite{2000SSRv...92..279B} has applied the method successfully to
simulations of low velocity collisions between brittle objects,
\cite{2004Icar..167..431S} used experimentally measured data to
parametrise the compressive strength curve of icy grain agglomerates
and derived a modified Murnaghan equation of state (see
sect.~\ref{sect:pm}) to simulate porous objects.

Recent astronomical investigations such as the Deep Impact mission
\citep{2006Sci...313..635L} and observations of comets (e.g.,
\citealt{2006A&A...458..669L}) yield strong indications for rather low
densities of comets (the values range from 0.1~$\mathrm{g/cm^3}$ to
several tenths of $\mathrm{g/cm^3}$ and still have some larger
errors). This matches fairly well with the idea of collisional growth of
fractal porous dust agglomerates to planetesimals.

In order to extend the investigations by \cite{2004Icar..167..431S},
we apply his model to simulate the collisions between equal sized
(ice) agglomerates with different impact parameters and also explore
the influence of material properties to the simulation outcome. The
outline of the paper is as follows: In the next section, we will
present the physical model, that is the basic equations and material
properties we assume. In sect.~\ref{sect:ni} we will discuss some
numerical issues and in sect.~\ref{sect:s} the setup of our
simulations.  We present our results in sect.~\ref{sect:r} and finally
draw some conclusions.

\section{Physical Model\label{sect:pm}}
\subsection{Basic Equations}

The system of partial differential equations that describe the
dynamics of an elastic solid body is given by three equations. The
first is the continuity equation
\begin{equation}
\frac{\mathrm{d}\varrho}{\mathrm{d}t} + \varrho \frac{\partial
v^\alpha}{\partial x^\alpha} = 0,
\end{equation}
where $\varrho$ denotes the density and $\mathbf{v}$ the velocity and where
the Einstein summation rule is applied. Greek indices denote the spatial
coordinates and run from 1 to 3. The second equation in the system
accounts for the conservation of momentum
\begin{equation}
\frac{\mathrm{d}v^\alpha}{\mathrm{d}t} = \frac{1}{\varrho}
\frac{\partial \sigma^{\alpha \beta}}{\partial x^\beta}.
\end{equation}
The stress tensor $\sigma$ is given by the pressure $p$ and the
deviatoric stress tensor $S^{\alpha\beta}$ according to
\begin{equation}
\sigma^{\alpha\beta} = -p\delta^{\alpha \beta} + S^{\alpha \beta}.
\end{equation}

In contrast to fluid dynamics, this set of partial differential
conservation equations is not sufficient to describe the elastic body,
since the time evolution of the deviatoric stress tensor is not yet
specified. The missing relations are called the constitutive equations
which describe the rheology of the body and relate the kinematic
states of the body to the dynamic states. The elastic behaviour of a
solid body can be described by Hooke's law, which reads in three
dimensions
\begin{equation}
S^{\alpha \beta} \sim 2\mu\left(\varepsilon^{\alpha \beta} - \frac{1}{3}
\delta^{\alpha \beta} \varepsilon^{\gamma \gamma}\right),
\end{equation}
where $\mu$ is the shear modulus of the material, and
$\varepsilon^{\alpha \beta}$ denotes the strain tensor which is given by
\begin{equation}
\varepsilon^{\alpha \beta} = \frac{1}{2} \left( \frac{\partial
x'^\alpha}{\partial x^\beta}  + \frac{\partial x'^\beta}{\partial
x^\alpha} \right).
\end{equation}
Here, the primed coordinates denote the coordinates of the deformed
body.

The stress rate has to be defined in a way that obeys the principle of
frame invariance. There are various possibilities to achieve this. We
follow the usual approach which is used for SPH codes
\citep{1994Icar..107...98B} and adopt the Jaumann rate form, where the
time evolution of the deviatoric stress tensor can be expressed as
\begin{equation}
\frac{\mathrm{d}S^{\alpha \beta}}{\mathrm{d}t} = 2\mu \left(
\dot{\varepsilon}^{\alpha\beta} - \frac{1}{3} \delta^{\alpha \beta}
\dot{\varepsilon}^{\gamma \gamma} \right) + S^{\alpha
\gamma}R^{\gamma\beta} +S^{\beta \gamma}R^{\gamma \alpha},
\end{equation}
where $R^{\alpha \beta}$ denotes the rotation rate tensor
\begin{equation}
\label{eq:rotation_rate_tensor}
R^{\alpha \beta} = \frac{1}{2} \left( \frac{\partial v^\alpha}{\partial
x^\beta} - \frac{\partial v^\beta}{\partial x^\alpha}\right),
\end{equation}
and $\dot{\varepsilon}^{\alpha \beta}$ the strain rate tensor
\begin{equation}
\dot{\varepsilon}^{\alpha \beta} = \frac{1}{2} \left( \frac{\partial v^\alpha}{\partial
x^\beta} + \frac{\partial v^\beta}{\partial x^\alpha}\right).
\label{eq:strain_rate_tensor}
\end{equation}

The closure of this set of equations is provided by the equation of
state that relates the pressure $p$ to the density of the agglomerate.
Here, we focus on porous objects and follow the semi-empirical approach
by \cite{2004Icar..167..431S}. We use an extension of the Murnaghan
equation of state which accounts for the change of porosity and reads
\begin{equation}
p(\varrho) = \left\{ \begin{array}{ll} 
	K(\varrho_0') (\varrho/\varrho_0' - 1), & \varrho < \varrho_\mathrm{c}, \\[1ex]
	\Sigma(\varrho), & \varrho \geq \varrho_\mathrm{c}.
	\end{array} \right.
	\label{eq:eos}
\end{equation}
The quantity $\varrho_0'$ is the reference density of an agglomerate at no
external stress while $\varrho_0$ is the initial density of the porous
agglomerate. Note that $\varrho_0'$ is in general different to the
initial density $\varrho_0$ of the porous agglomerate after the start of
the simulation. The critical density $\varrho_\mathrm{c}$ is a function
of the reference density $\varrho_0'$ and determines the state where the
pressure reaches the compressive strength limit $\Sigma(\varrho)$. As
soon as the pressure decreases, the behaviour of the agglomerate is
again elastic but with a different bulk modulus $K(\varrho_0')$ since the
reference density has changed. The slope of $K(\varrho_0')$ has to be
known either by experimental data or theoretical considerations.
The tensile strength of the material is determined accordingly and
limits the tension for negative pressure.

The sound speed of the agglomerate is given by the bulk
modulus of the agglomerate and the reference density according to the relation
\begin{equation}
c(\varrho_0') = \sqrt{ K(\varrho_0')/\varrho_0'}.
\end{equation}
For plastic states when the pressure exceeds the compressive strength
of the material the sound speed is calculated according to 
\begin{equation}
c(\varrho) = \sqrt{\frac{\mathrm{d} \Sigma (\varrho)}{\mathrm{d}
\varrho}}.
\end{equation}

\cite{2004Icar..167..431S} additionally uses an empirical damage model
for porous agglomerates which is based on the damage model for brittle
materials developed for SPH by \cite{1994Icar..107...98B,benz:1995}.
However, this model is only applicable for the simulation
of brittle fracture in rocks \citep{grady:1980}, and our test simulations
show that the application of the damage model to porous agglomerates
does not yield reliable results since the model also includes
compressive damage effects. Because of these considerations, we do not
include any damage model in our calculations and only consider the
fragmentation due to plastic flow.

\subsection{Material properties}
The shape of the compressive and tensile strength curves have to be
known either by experimental data or theoretical considerations. Since
we want to compare our calculations to \cite{2004Icar..167..431S}, we
primarily use the values published in his paper which are derived from
experimental data (see \citealt{kendall:1987,valverde:1998}): The
compressive strength is given by $\Sigma(\varphi) = \Sigma_0
\varphi^\alpha$, with $\varphi=\varrho/\varrho_0$, and the tensile
strength accordingly by $T(\varphi) = T_0 \varphi^\beta$. The
dependence of the bulk modulus of the aggregate on the porosity reads
$K(\varphi)=K_0\varphi^\gamma$. The shear modulus is assumed to be
$\mu = K/2$, and the shear strength is defined by $Y(\varphi) =
\sqrt{2T(\varphi)\Sigma(\varphi)/3}$. The parameter $\varphi$ is
related to the porosity $\phi=\varrho/\varrho_\mathrm{s}$ in the
following way
\begin{equation}
\varphi = \frac{\varrho}{\varrho_0} = \phi
\frac{\varrho_\mathrm{s}}{\varrho_0} = \frac{\phi}{\phi_0},
\end{equation}
where $\varrho_\mathrm{s}$ denotes the density of the solid material in
the porous body and $p_0$ is the porosity of the body at the initial
simulation time.

The transition from an elastic state to a yield state of a material can
be characterised by a complex stress state. The stress at which the
transition starts is called the yield stress and the condition is called
the yield criterion or the plasticity criterion. In this paper, we
follow the implementation by \cite{benz:1995} and assume isotropic material
and apply the von Mises yield criterion: We calculate the
second irreducible invariant of the deviatoric stress tensor
$J_2=S_{\alpha \beta}S^{\alpha\beta}$ and reduce the deviatoric stress
when necessary according to
\begin{equation} S^{\alpha\beta} \rightarrow f S^{\alpha\beta}, \qquad f
= \min\left[ Y(\varphi)^2/3J_2, 1\right]. \end{equation}

\section{\label{sect:ni}Numerical Issues}
Smooth Particle Hydrodynamics (SPH) is perfectly suitable for the
simulation of brittle and plastic materials. The continuum of the solid
body is discretised into mass packages which are called particles. The
particles move like point masses according to the Lagrangian form of the
equation of motion. They carry all physical properties like mass, momentum
and energy of the part of the solid body they represent. The
particles interact by kernel interpolation during the simulation to
exchange momentum and energy. For a complete description of the method
and its features and qualities, we refer to two comprehensive review
articles \citep{benz:1990,monaghan:1992}.

In standard SPH, the velocity derivatives in
eqs.~(\ref{eq:rotation_rate_tensor}) and (\ref{eq:strain_rate_tensor}) for the determination of the rotation rate and the strain
rate tensor for particle $i$ are usually calculated according to
\begin{equation}
\frac{\partial v_i^\alpha}{\partial x_i^\beta} = \sum_j
\frac{m_j}{\varrho_j} (v^\alpha_j - v^\alpha_i) \frac{\partial
W^{ij}}{\partial x^\beta_i},
\end{equation}
where the sum runs over all interaction partners $j$ of particle $i$,
and $W^{ij}$ denotes the kernel for the particular interaction.

This approach, however, leads to erroneous results and does not conserve
angular momentum due to the discretisation error by particle disorder in
simulations of solid bodies. This error can be avoided by constructing
and applying a correction tensor $C^{\gamma \beta}$ according to 
\begin{equation}
\frac{\partial v_i^\alpha}{\partial x_i^\beta} = \sum_j
\frac{m_j}{\varrho_j} (v^\alpha_j - v^\alpha_i) \sum_\gamma \frac{\partial
W^{ij}}{\partial x^\gamma_i}C^{\gamma \beta},
\label{eq:ext_sph}
\end{equation}
where the correction tensor $C^{\gamma \beta}$ is the inverse of
\begin{equation}
\sum_j \frac{m_j}{\varrho_j} (x^\alpha_j - x^\alpha_i) \frac{\partial
W^{ij}}{\partial x^\gamma_i},
\end{equation}
that is
\begin{equation}
\sum_j \frac{m_j}{\varrho_j} (x^\alpha_j - x^\alpha_i) \sum_\gamma
\frac{\partial W^{ij}}{\partial x^\gamma_i} C^{\gamma \beta} = \delta^{\alpha
\beta}.
\end{equation}
By applying this correction tensor first order consistency can
be constructed where the errors due to particle disorder cancel out
and the conservation of angular momentum is ensured. This is
similar to an approach that \citet{bonet:1999a} proposed for the
conservation of angular momentum, where all spatial derivatives are
corrected according to eq.~(\ref{eq:ext_sph}). We have found however that
it is sufficient to correct only the rotation rate and the strain
rate tensor.

We additionally use the standard SPH artificial viscosity
\citep{monaghan:1983} with $\alpha=1$. However, in order to reduce the
viscous energy dissipation, we follow Sirono's approach and apply the
artificial viscosity only when the approaching relative velocity of
two particles is larger than the sound speed in contrast to the
standard case where the artificial viscosity acts always on
approaching particles.

Our SPH code {\tt parasph} \citep{hipp:2004} has been successfully
tested with some numerical standard tests such as the collision of
perfectly elastic rings \citep{swegle:1992, monaghan:2000} and the
simulations of ductile and brittle rods under tension
\citep{1994Icar..107...98B,gray:2001}. Additionally, the Sirono model
in the extended code has been checked carefully \citep{schaefer2005}
and we have successfully reproduced the impact result for the standard
setup in \cite{2004Icar..167..431S}.

\section{\label{sect:s}Setup}

All collisions have been simulated in three dimensions.  The setup of
our calculations includes the following parameters: Two spherical ice
agglomerates with the initial density $\varrho_0=0.1~\mathrm{g/cm^3}$
and identical radius $r=1~\mathrm{m}$ collide at either
$20~\mathrm{m/s}$ or $10~\mathrm{m/s}$. The initial density implies
an initial porosity of 90~\%. The density dependencies of compressive
and tensile strengths are chosen according to
\cite{2004Icar..167..431S} to $\Sigma(\varrho)=\Sigma_0
(\varrho/\varrho_0)^6$ and $T(\varrho) = T_0 (\varrho/\varrho_0)^5$,
and the bulk modulus is given by $K(\varrho) =
K_0(\varrho/\varrho_0)^4$. We always assume the shear modulus is given
by $\mu=K/2$. For the first simulations with impact velocity
$20~\mathrm{m/s}$ we use the basic setup values from Sirono and set
$K_0=6\times 10^5~\mathrm{Pa}$, $\Sigma_0=600~\mathrm{Pa}$, and
$T_0=6\times10^3~\mathrm{Pa}$. These are the values for which Sirono
(and we) find perfect sticking behaviour for a smaller sphere
impacting into a larger object (with a radii ratio of 10:3) at speed
$10~\mathrm{m/s}$.  The temperature implicitely assumed in this set of
parameters corresponds to very cold ice, which is consistent with the
simulation results where large deformations but hardly any compression
is found. Since the probability of a head-on collision is low, we
have additionally varied the impact parameter of the collision,
$b=0.2$, $0.4,$ $0.8$, $1.2~\mathrm{m}$. The results of these
simulations are shown in sect.~\ref{sect:impact}.

To study the importance of realistic material properties, we have then
varied the tensile and compressive strength parameters $T_0$ and
$\Sigma_0$ in some of the $10~\mathrm{m/s}$ head-on collisions and
present these calculations in sect.~\ref{sect:strength}. Reducing the
impact velocity also allows to compare the influence of the latter.

\section{\label{sect:r}Results}

\subsection{\label{sect:impact}Varying impact parameter}
\begin{figure} \resizebox{\hsize}{!}{
\includegraphics{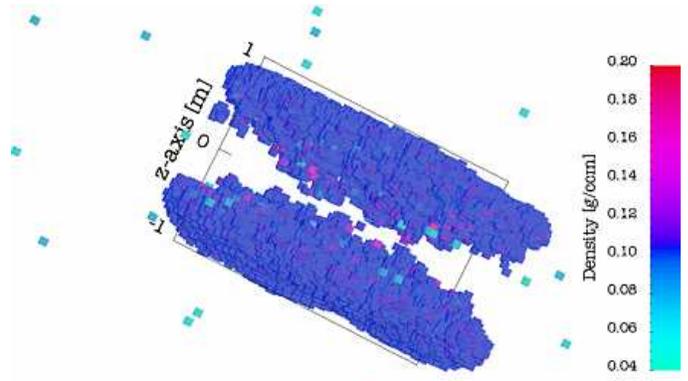} }
\caption{\label{fig:head-on}Head-on collision with impact velocity
$20~\mathrm{m/s}$. The material strength coefficients are chosen to
$\Sigma_0=600~\mathrm{Pa}$ and $T_0=6000~\mathrm{Pa}$. The plot shows
colour-scaled the density for each particle at simulation time
$t=0.662~\mathrm{s}$ after contact. The colourbar is in units of
$\mathrm{g/cm^3}$ and is valid for all colour-scale plots in this
paper. Also note that the line of approach is tilted by $\sim 30\degr$
from the vertical in all plots.}
\end{figure}

\begin{figure} \resizebox{\hsize}{!}{
\includegraphics{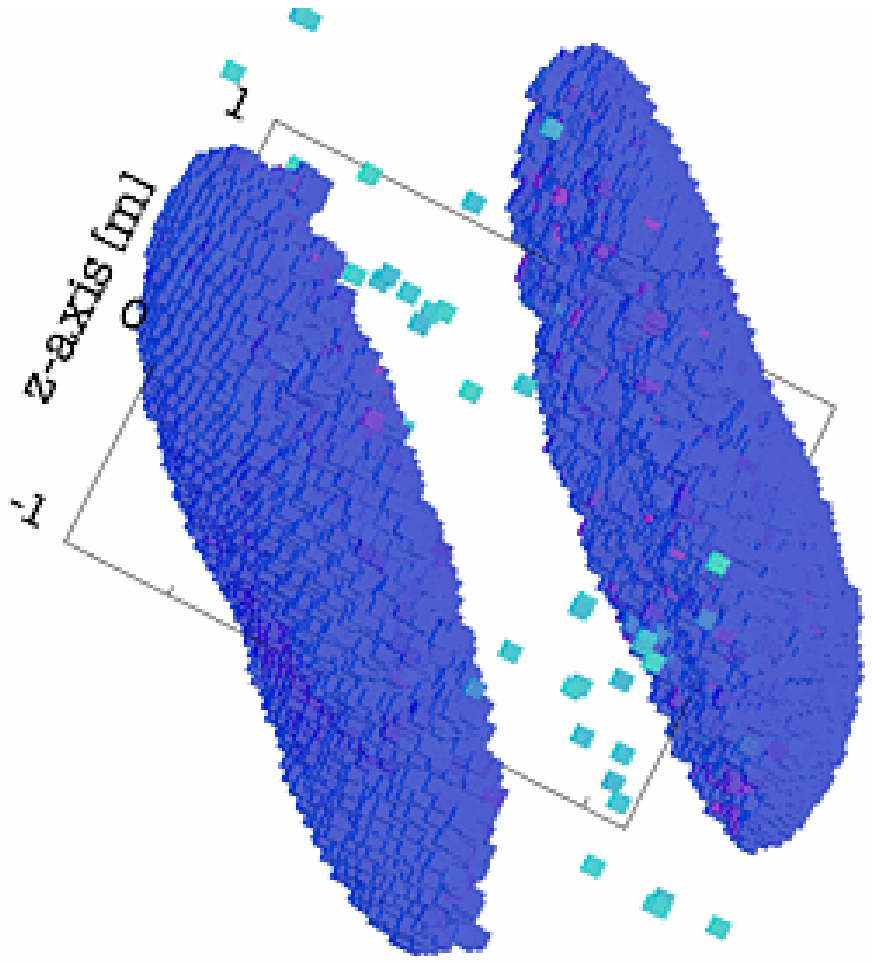} }
\caption{\label{fig:b0.2}Collision with impact parameter
$b=0.2~\mathrm{m}$.  The plot shows colour-scaled the density
for each particle at simulation time $t=0.662~\mathrm{s}$ after contact.}
\end{figure}

\begin{figure} \resizebox{\hsize}{!}{
\includegraphics{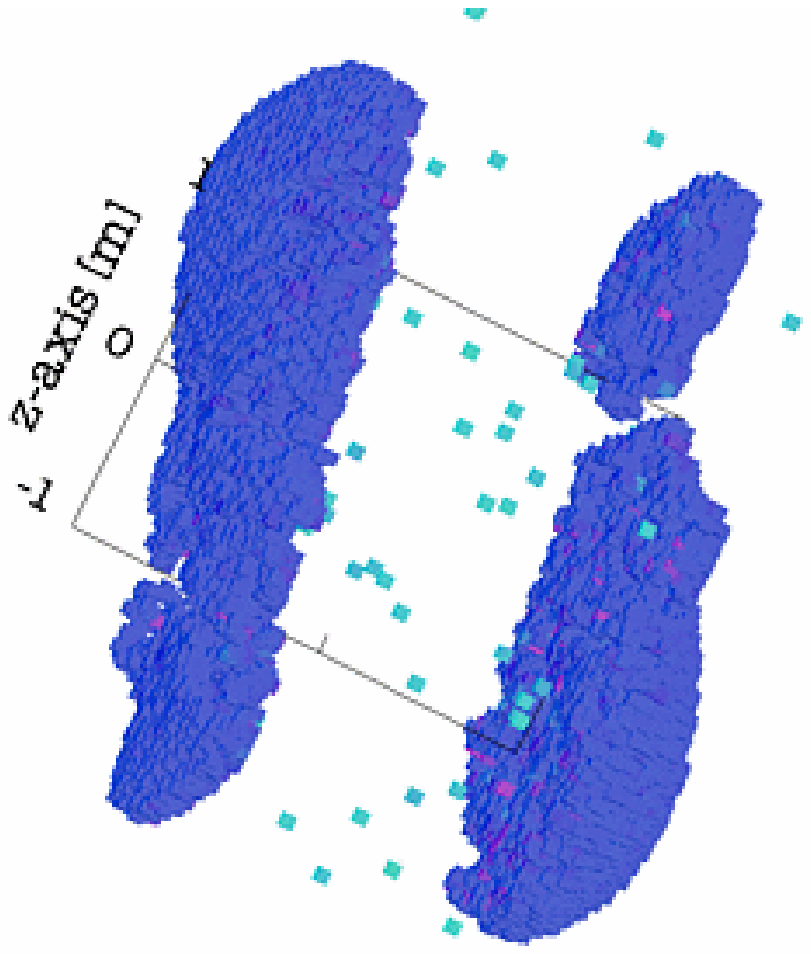} }
\caption{\label{fig:b0.4}Collision with impact parameter
$b=0.4~\mathrm{m}$. The plot shows colour-scaled the density for
each particle at simulation time $t=0.662~\mathrm{s}$ after contact. } 
\end{figure}

\begin{figure} \resizebox{\hsize}{!}{
\includegraphics{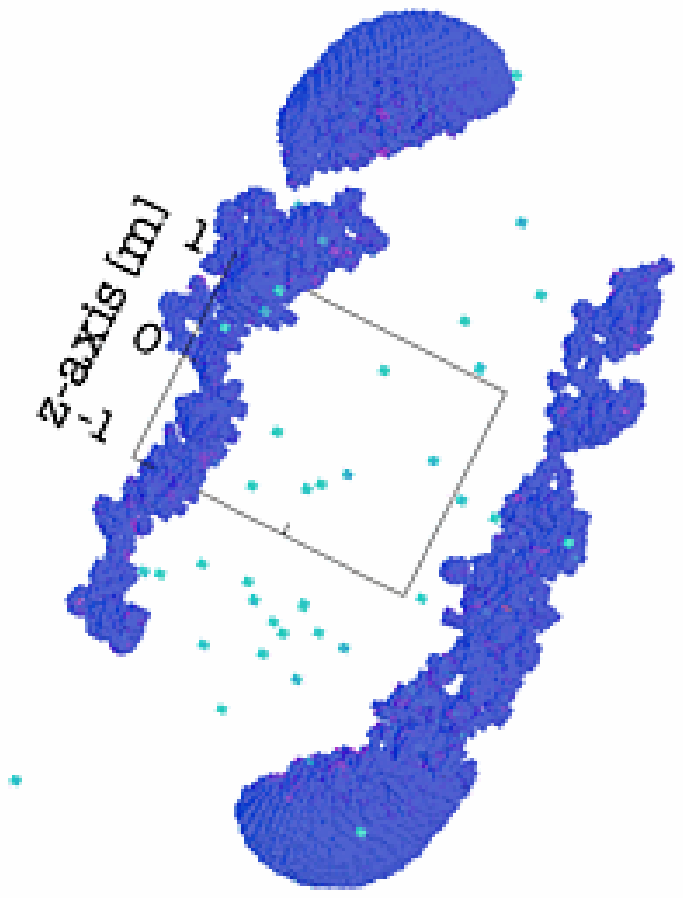} }
\caption{\label{fig:b0.8}Collision with impact parameter
$b=0.8~\mathrm{m}$.  The plot shows colour-scaled the density
for each particle at simulation time $t=0.662~\mathrm{s}$ after contact.  }
\end{figure}

\begin{figure} \resizebox{\hsize}{!}{
\includegraphics{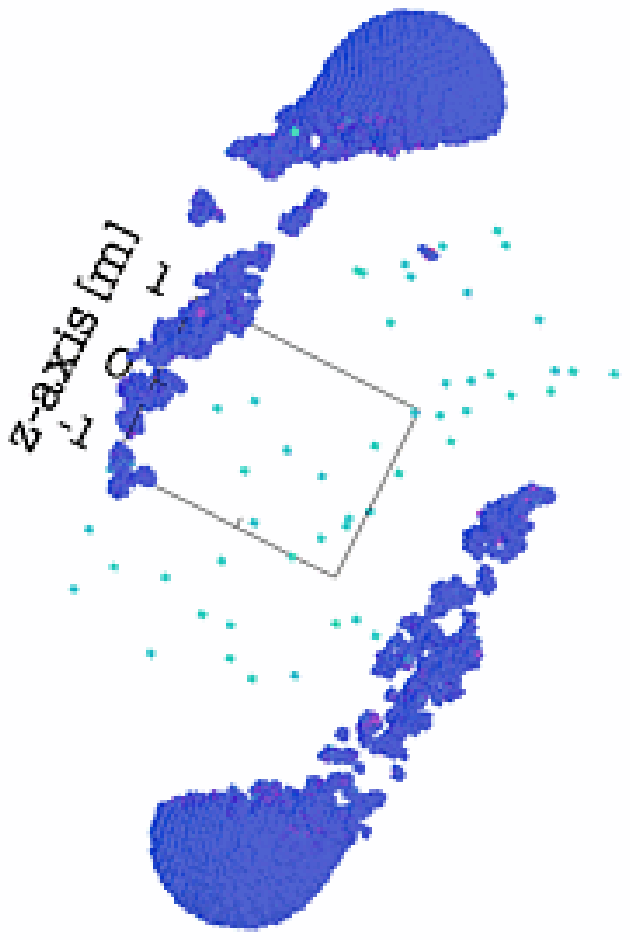} }
\caption{\label{fig:b1.2}Collision with impact parameter
$b=1.2~\mathrm{m}$.  The plot shows colour-scaled the density
for each particle at simulation time $t=0.662~\mathrm{s}$ after contact.  }
\end{figure}

As noted above, in this section we focus on the material parameters
for which \cite{2004Icar..167..431S} found perfect sticking for a
collision with a projectile 3:10 smaller in radius than the target.
Figs.~\ref{fig:head-on}-\ref{fig:b1.2} present the simulation outcome
for five different impact parameters. The plots show the SPH particles
with a colour-scale code\footnote{As in the printed version of this
paper the figures have to be displayed in grey-scale, we refer the
reader to the online-version for colour figures.} for the density of
each particle.  As we are mainly interested in the final configuration
of the impact after compression has ceased, the density is acting as a
measure for any permanent compaction that may have taken place during
the compression phase.

Each sphere consists of approximately 11\,500 SPH particles which are
initially placed on a tetrahedral grid to maximise the number of
adjacent and interacting particles and thus to increase the
resolution. The figures show the surface of the resulting objects
after impact, as the particles are plotted opaque.

The first simulation (see fig.~\ref{fig:head-on}) demonstrates the
behaviour during a head-on collision with impact parameter $b=0$. In
contrast to the sticking mechanism found by \cite{2004Icar..167..431S}
with the same material properties but different sizes, the two spheres
do not penetrate. In fact, they undergo a large change in shape and
the two spheres end up as flattened discs. Almost the entire kinetic
energy of the impact is dissipated by plastic deformation, and
ultimately the two discs are at rest in the barycentric system. Only
some particles which were tossed outwards during the compression phase
fly away at constant speed.

The picture slightly changes for the simulation with $b=0.2~\mathrm{m}$
(see fig.~\ref{fig:b0.2}).
As in the head-on collision, the two spheres transform into flattened
discs with some separate particles floating away. In contrast to the
head-on collision, the remnants of the spheres rotate slowly.

First major difference arises at the step from $b=0.2~\mathrm{m}$ to
$b=0.4~\mathrm{m}$ (see fig.~\ref{fig:b0.4}), when the spheres break up.
Again, the spheres undergo a large plastic deformation during the
compression phase and become strongly prolate. A small piece is quarried
out of each object after the compression due to the strong
tensile forces caused by rotation.

Even more smaller fragments emerge after the collision with impact
parameter $b=0.8~\mathrm{m}$ (see fig.~\ref{fig:b0.8}). However, now
the largest remnants of the collision consist of two rotating
half-spheres. The density of the half-spheres is exactly the density
of the initial spheres. Essentially they have not been compacted.

The situation hardly changes for $b=1.2~\mathrm{m}$ (see
fig.~\ref{fig:b1.2}). Again, smaller fragments are tossed out from the
spheres after the compaction phase.  The two formed larger remnants
have again the shape of half-spheres, slightly larger than in the
simulation with $b=0.8~\mathrm{m}$. The rotation rate and the velocity
of the half-spheres is larger than for smaller impact parameters.

\begin{figure}
\resizebox{\hsize}{!}{
\includegraphics{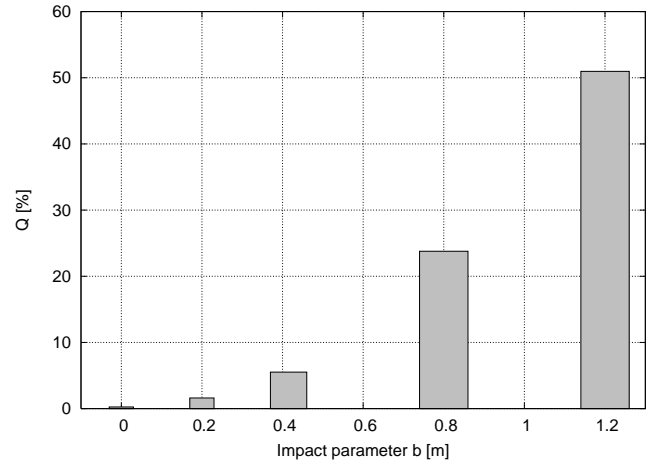}
}
\caption{\label{fig:ekin}Ratio $Q$ between kinetic energy at end of the
simulation and initial energy for five different impact
parameters.}
\end{figure}

The strong influence of the impact parameter $b$ on the change in
kinetic energy with the respect to the centre of mass is shown in
fig.~\ref{fig:ekin}, where the ratio $Q$ of the kinetic energy at
the end of the simulation to the initial kinetic energy is plotted. In
the case of the head-on collision, more than 99.7~\% of the initial
kinetic energy has been dissipated during the plastic deformation phase. The
remnants (except for a few particles) of the collision are at rest.
Although they do not stick together, their relative velocity is zero.
Even for impact parameters $b=0.2$ and $0.4$~$\mathrm{m}$, the kinetic
energy at the end is less than 10~\% of the initial kinetic energy. The
spheres are effectively slowed down during the compression phase. 

The situation changes for higher impact parameters. For
$b=0.8~\mathrm{m}$ less than 80~\% and for $b=1.2~\mathrm{m}$ even less
than 50~\% of the initial kinetic energy is lost due to the compression.

Interestingly, the density of the remnants after the simulation is
basically the initial density for all studied impact parameters. Only
some small areas in the contact region have a slightly higher density.
The density in the interiour of the objects does not differ distinctly
from the density at the surfaces. Therefore only the latter is shown
in the figures.

\subsection{\label{sect:strength}Varying material strength}

Although \cite{2004Icar..167..431S} has investigated the influence
of the material strength on the collisional outcome, we have
additionally studied the behaviour for our equal-sized agglomerates.

\begin{figure} \resizebox{\hsize}{!}{
\includegraphics{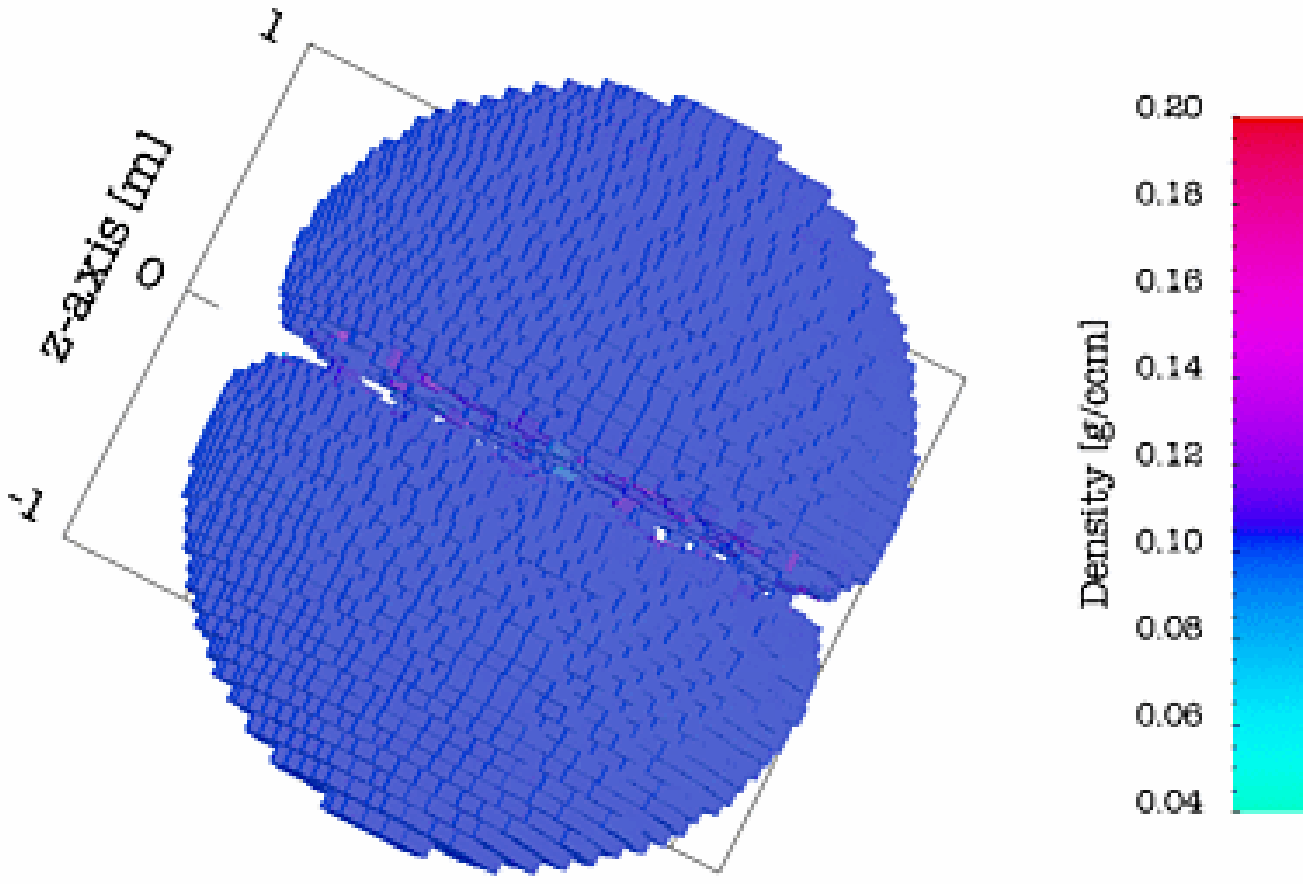} }
\caption{\label{fig:6006000} Head-on collision with impact velocity
$10~\mathrm{m/s}$. The material strength coefficients are
$\Sigma_0=600~\mathrm{Pa}$ and $T_0=6\,000~\mathrm{Pa}$. The plot
shows colour-scaled the density for each particle at simulation time
$t=0.5~\mathrm{s}$ after contact.}  
\end{figure}

\begin{figure} \resizebox{\hsize}{!}{
\includegraphics{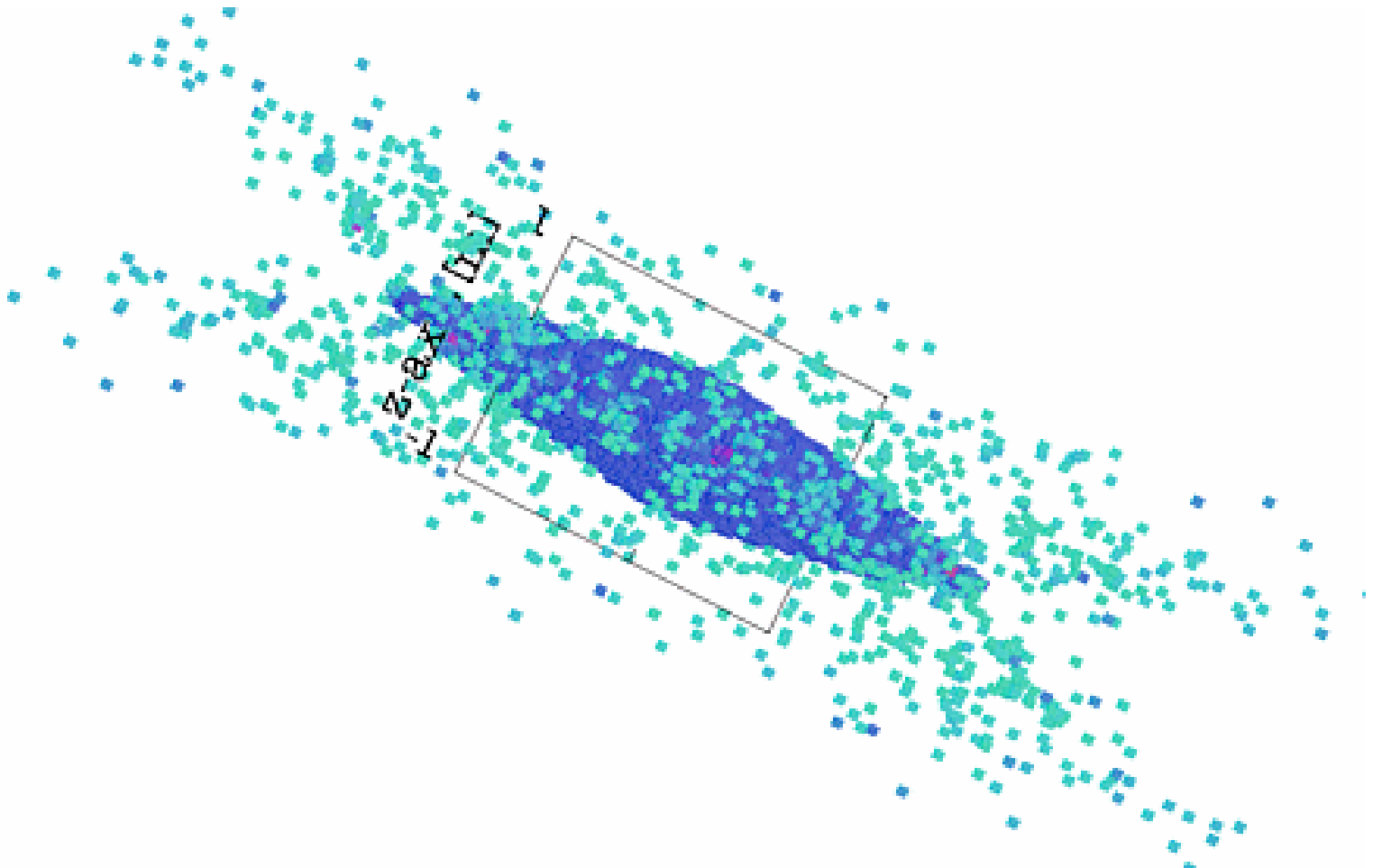} }
\caption{\label{fig:600600} Head-on collision with impact velocity
$10~\mathrm{m/s}$. The material strength coefficients are
$\Sigma_0=600~\mathrm{Pa}$ and $T_0=600~\mathrm{Pa}$. The plot shows
colour-scaled the density for each particle at simulation time
$t=0.5~\mathrm{s}$ after contact.}
\end{figure}

\begin{figure} \resizebox{\hsize}{!}{
\includegraphics{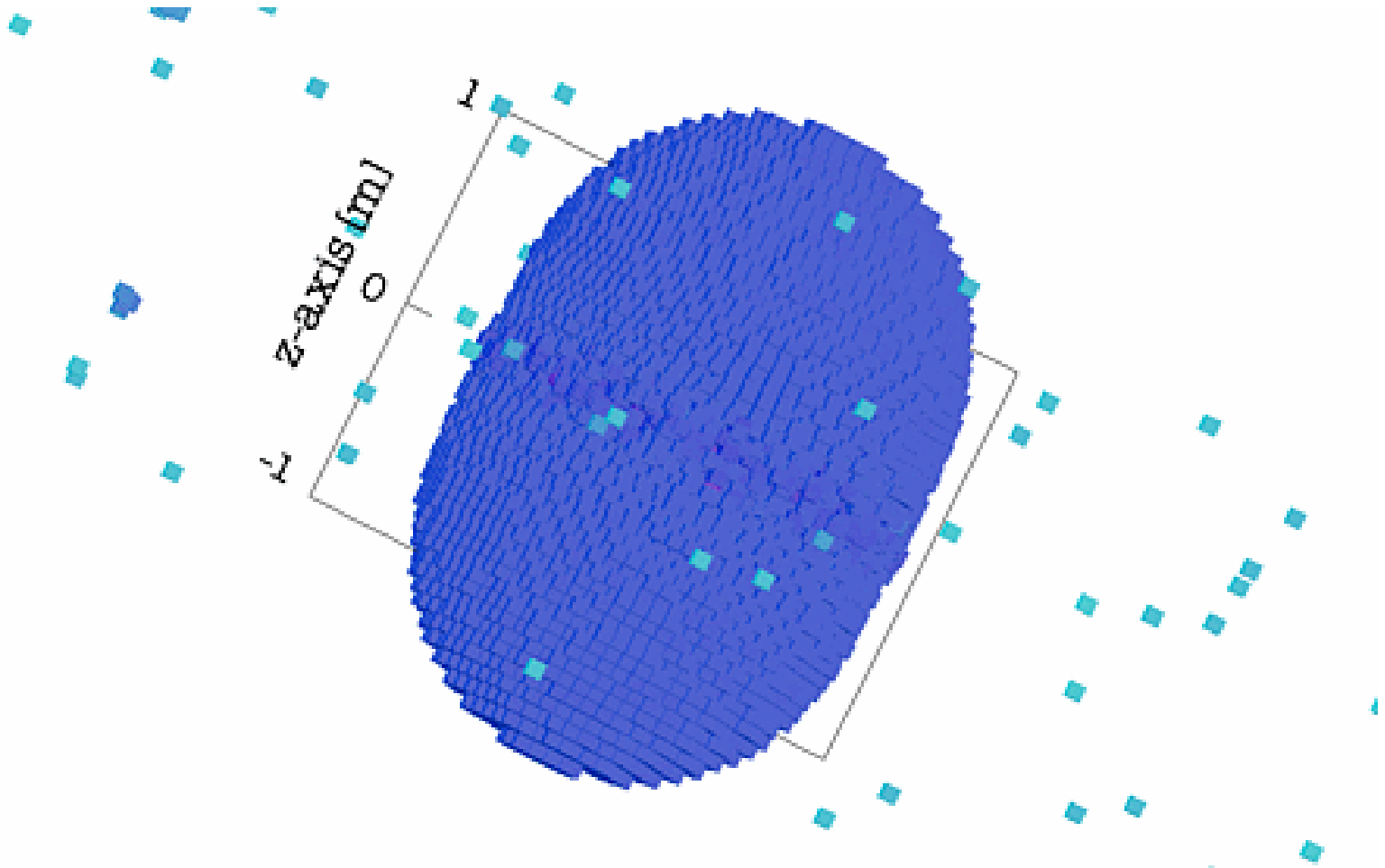} }
\caption{\label{fig:60006000} Head-on collision with impact velocity
$10~\mathrm{m/s}$. The material strength coefficients are
$\Sigma_0=6\,000~\mathrm{Pa}$ and $T_0=6\,000~\mathrm{Pa}$. The plot
shows colour-scaled the density for each particle at simulation time
$t=0.5~\mathrm{s}$ after contact.}
\end{figure}

\begin{figure} \resizebox{\hsize}{!}{
\includegraphics{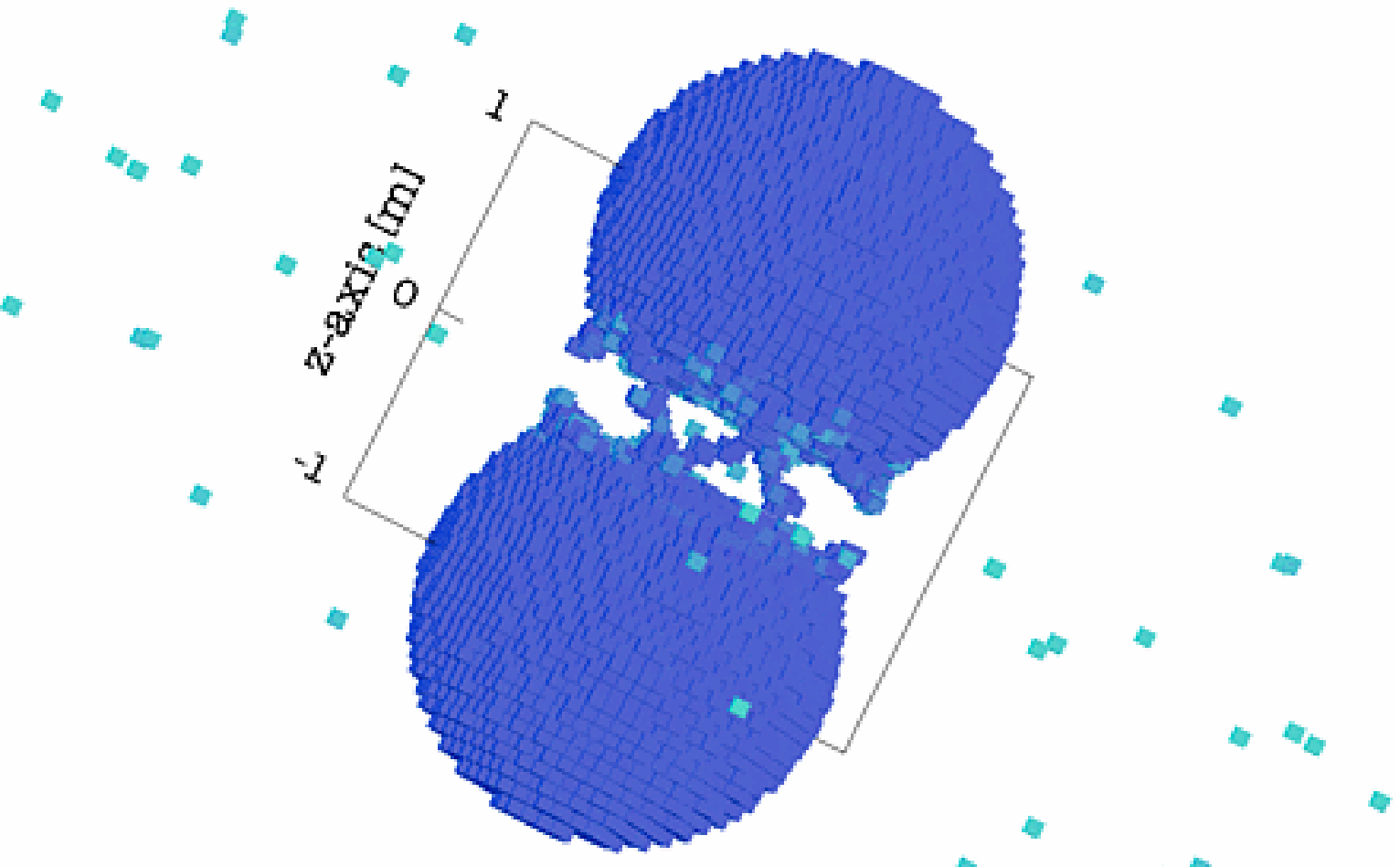} }
\caption{\label{fig:600006000} Head-on collision with impact velocity
$10~\mathrm{m/s}$. The material strength coefficients are
$\Sigma_0=60\,000~\mathrm{Pa}$ and $T_0=6\,000~\mathrm{Pa}$. The plot
shows colour-scaled the density for each particle at simulation time
$t=0.5~\mathrm{s}$ after contact.}
\end{figure}

\begin{figure} \resizebox{\hsize}{!}{
\includegraphics{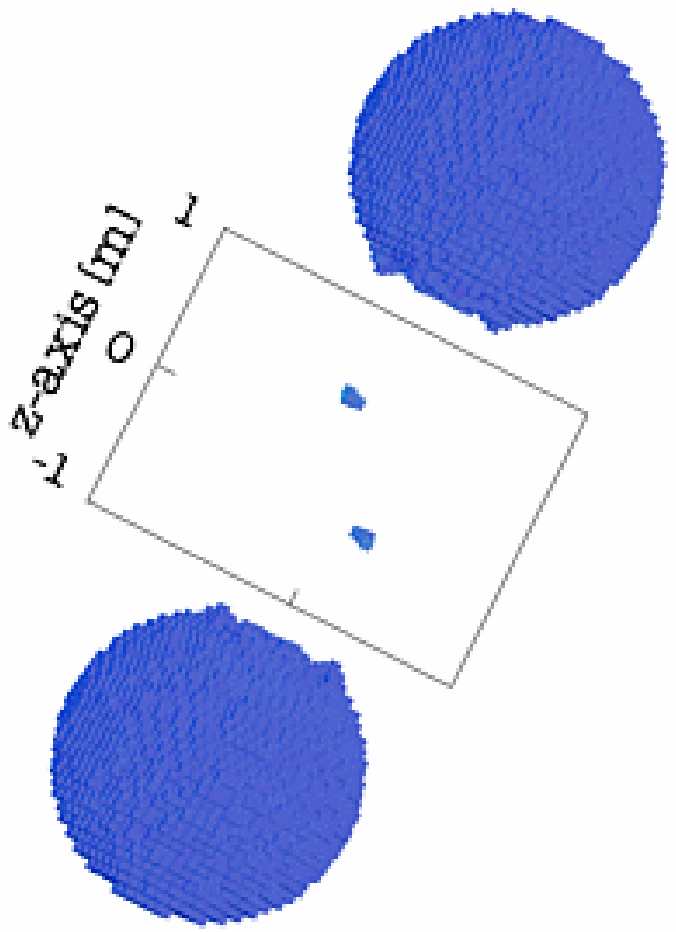} }
\caption{\label{fig:6000060000} Head-on collision with collision
velocity $10~\mathrm{m/s}$. The material strength coefficients are
$\Sigma_0=60\,000~\mathrm{Pa}$ and $T_0=60\,000~\mathrm{Pa}$. The plot
shows a colour-scaled the density for each particle at simulation
time $t=0.5~\mathrm{s}$ after contact.} \end{figure}

All simulations with varying material strengths are head-on collisions
with a relative velocity of $10~\mathrm{m/s}$. The basic relations for
the compressive and tensile strengths and the bulk modulus remain
unchanged, only the coefficients $\Sigma_0$ and $T_0$ are varied. 

The figures \ref{fig:6006000}-\ref{fig:600006000} show colour-scaled
density plots of the particles at the end of each simulation. For the
first of these simulations, we have used the same material properties
as in sect.~\ref{sect:impact}, $\Sigma_0=600$~$\mathrm{Pa}$ and
$T_0=6\times 10^3~\mathrm{Pa}$ (see fig.~\ref{fig:6006000}). Because
the collision velocity is half the collision velocity from the
simulation shown in fig.~\ref{fig:head-on}, the spheres do not
experience the same drastic plastic deformation. In fact, they end up
as two solid half spheres and no particles were pulled out of the
spheres during the compression phase. Their relative velocity at the
end are again vanishing. This picture changes drastically if the
tensile strength coefficient $T_0$ is reduced by one order of
magnitude to $600~\mathrm{Pa}$ (see fig.~\ref{fig:600600}). Now the
tensile forces are too weak to sustain the bodies during the contact
phase. In the end, the two spheres stick together into a disc-like
structure and many particles are ejected due to the fragmentation by
plastic flow. If the compressive strength parameter $\Sigma_0$ is
increased to $6\times10^3~\mathrm{Pa}=T_0$ instead (see
fig.~\ref{fig:60006000}), the material is strong enough to evade the
fragmentation, only a few particles are thrown out, and the two
spheres finally stick together, forming an elongated
body. Interestingly, if the compressive strength is augmented even
further to $6\times10^4~\mathrm{Pa}$ (see fig.~\ref{fig:600006000}),
the bodies become too elastic and do not stick. The basic structure of
the spheres is conserved, only in the impact contact region, the
spheres are flattened. The objects rather bounce off from each other,
losing some particles. A nearly fully elastic rebound of the spheres
happens for $\Sigma_0=T_0=6\times10^4~\mathrm{Pa}$ (see
fig.~\ref{fig:6000060000}).

Again, the density of the spheres does not change significantly. Although
their shapes change notably, they are not compacted effectively and
their porosities stay mainly unchanged.

\section{Discussion} 

The simulations presented in this work extend the investigation of
\cite{2004Icar..167..431S} to collisions between equal-sized
agglomerates. Sirono's main results regarding collisional growth and
stickiness conditions can be summarised in the following way: Porous
agglomerates can grow by collisions if the tensile strength is larger
than the compressive strength, the shear strength is larger than the
compressive strength and the Mach number is lower than $0.04$ for
oblique impacts, which corresponds in our simulations to a maximum
collision velocity of $3.1~\mathrm{m/s}$. Additionally, a damage
restoration effect has to be included, otherwise the agglomerates are
totally fragmented. Since we do not use the damage model from Sirono,
there is no need for a restoration effect.

Our new simulations with varying impact parameters indicate that collisions
with $b<0.4$ lead to drastic shape deformations, the spheres finally
form disc-like structures. Clearly, disc-like structures will withstand
subsequent collisions even worse if the impact is perpendicular to the
disc plane. On the other hand most (for head-on collisions nearly all)
of the kinetic energy is lost, which possibly leads to lower collision
velocities for following encounters, thus favouring growth in subsequent
collisions. The change in shape is lower for larger impact parameters
$b>0.4$, where less kinetic energy is lost. However, agglomerates emerging
from oblique collisions tend to rotate and eventually break if the
tensile strength is not large enough. Then, plastic flow leads to
fragmentation of agglomerates after the compression phase. However, the
tensile strength of a realistic porous protoplanetesimal might be higher
in orders of magnitude than assumed here.

One of the most striking simulation results is the missing compaction
of the agglomerates. Although they are rather porous objects (with
90~\% porosity), their porosity does not change significantly during
the impact. It is unclear, how the porosity in the agglomerates may be
reduced and the bodies compacted. Probably, lower collision velocities
and differing sizes of the colliding agglomerates lead to the
incorporation of smaller bodies into large agglomerates, and a slight
decrease of porosity. However, the velocity distribution of
protoplanetesimals in the solar nebula at the Earth's orbital distance
ranges up to several tens of $\mathrm{m/s}$.  One collision at such a
high speed might destroy previous growth entirely.

The varying material strength parameters which were investigated in the
second series of simulations show the strong influence of the material
properties on the collisional outcome. For the applied values we find the
whole spectrum: sticking, no sticking but vanishing relative velocity,
complete fragmentation, and elastic rebound. In contrast to Sirono's
results, the only simulation that results in sticking uses
$\Sigma_0=T_0>Y_0$. Moreover, since the Mach number of our simulations
is $0.13$ for the $10~\mathrm{m/s}$ collisions and $0.26$ for the
$20~\mathrm{m/s}$ collisions respectively, all simulations should lead
to fragmentation according to Sirono's results. 

Altogether the results indicate that to prevent fragmentation, a large
compressive strength may compensate a low tensile strength (see
fig.~\ref{fig:600006000}) and vice versa (see fig.~\ref{fig:6006000}).

\section{Conclusion}

As long as we do not have more insight into realistic material
properties of protoplanetesimals, it is cumbersome to give any
accurate description about the formation of planetesimals by
collisional growth. However, we have demonstrated that specific
strength parameters can lead to growth and encourage the formation of
planetesimals even for collisions of equal sized objects, where
previous investigations have found destruction.  Currently, it is a
major topic to establish realistic material parameters of
protoplanetesimals in laboratory experiments.  Recent experimental
data
\citep{2004PhRvL..93k5503B,2005PhRvE..71b1304W,2005Icar..178..253W}
provide first insight in the properties of porous dust agglomerates
which differ significantly from the properties applied in this study
and may lead to further perception. We plan to recalibrate the
existing SPH-model which was applied in this paper by performing
comparison calculations of impact experiments. Thus, a realistic model
for elastic and plastic behaviour of porous agglomerates will be
developed.

Additionally, our studies indicate that rotation is an important
unexplored effect which needs to be probed in more detail. A
subsequent study will therefore focus on the collisions between
rotating porous protoplanetesimals.

\begin{acknowledgements}
Part of this work was funded by the German Science Foundation (DFG) in
the frame of the Collaborative Research Centre SFB~382 {\em Methods
and Algorithms for the Simulation of Physical Processes on
Supercomputers}.  Additionally, CS has been supported by the DFG
through grant {\em KL 650/2}. We want to thank Michael Hipp for the
beautiful parallel SPH Code {\tt parasph} and useful comments for the
applied extensions and J{\"u}rgen Blum and Gerhard Wurm for fruitful
discussions. Additionally, CS thanks Sin-iti Sirono for helpful
comments during the preparation of this paper.  We also thank the
referee Lindsey Chambers for her clarifying remarks.
\end{acknowledgements}

\bibliographystyle{aa}
\bibliography{7354}

\end{document}